\documentclass[aps,prb,twocolumn,showpacs]{revtex4}
\usepackage{amssymb}

\usepackage{graphicx}

\begin{document}
\title{Strong and nonmonotonic temperature dependence of Hall coefficient in superconducting K$_x$Fe$_{2-y}$Se$_2$ single crystals}

\author{Xiaxin Ding, Yiming Pan, Huan Yang and Hai-Hu Wen}\email{hhwen@nju.edu.cn}

\affiliation{Center for Superconducting Physics and Materials,
National Laboratory of Solid State Microstructures and Department
of Physics, Nanjing University, Nanjing 210093, China}

\begin{abstract}
In-plane resistivity, magnetoresistance and Hall effect
measurements have been conducted on quenched K$_x$Fe$_{2-y}$Se$_2$
single crystals in order to analysis the normal-state transport
properties. It is found that the Kohler's rule is well obeyed
below about 80 K, but clearly violated above 80 K. Measurements of the Hall coefficient reveal a
strong but non-monotonic temperature dependence with a maximum at about 80 K, in contrast to
any other FeAs-based superconductors. With the two-band model
analysis on the Hall coefficient, we conclude that a gap may open
below 65 K. The data above 65 K are interpreted as a temperature
induced crossover from a metallic state at a low temperature to an
orbital-selective Mott phase at a high temperature. This is
consistent with the recent data of angle resolved photoemission
spectroscopy. These results call for a refined theoretical
understanding, especially when the hole pockets are absent or
become trivial in K$_x$Fe$_{2-y}$Se$_2$ superconductors.
\end{abstract}

 \maketitle

\section{Introduction}
In iron-based superconductors (FeSCs), it is very important to understand the electron correlation in different bands and the orbital selective Mott transitions. As revealed by angle
resolved photoemission spectroscopy (ARPES) and band
structure calculations, the superconductivity and normal state of
FeSCs are governed by their electronic structure involving the Fe
3d orbitals crossing the Fermi energy\cite{hirschfeld2011gap}. Due
to different structures and probably different methods for
fabrication, the iron chalcogenide superconductors show up with
different superconducting transition temperatures, for example,
$T_c \approx$ 8 K for Fe$_{1-x}$Se single
crystals\cite{hsu2008superconductivity}, $T_c \approx$ 16 K for
FeSe$_{1-x}$Te$_x$\cite{yeh2008tellurium}, $T_c \approx$ 32 K for
K$_x$Fe$_{2-y}$Se$_2$\cite{guo2010superconductivity, fang2011fe}. Furthermore,
a superconducting-like gaped feature can even be found in the
monolayer thin film of FeSe/SrTiO$_3$ at 65
K\cite{liu2012electronic,he2013phase}, this brings about new
vitality in exploring high temperature superconductivity in the
iron pnictide/chalcogenide systems.

In 2010, the discovery of the new K$_x$Fe$_{2-y}$Se$_2$
superconductor has generated great excitement in the
community\cite{wen2012overview}. Previously, it was proposed that
the pair-scattering of electrons between the hole and electron pockets
might drive the electron pairing with an $S^\pm$ pairing
symmetry\cite{mazin2008unconventional,kuroki2008unconventional}.
Nevertheless, the superconducting K$_x$Fe$_{2-y}$Se$_2$ seems
containing only electron pockets around the M-point, the hole
pockets which appear in most FeAs based superconductors near the $\Gamma$-point
are absent\cite{shein2011electronic}, which certainly leads to a
challenge to explain the superconductivity by the nesting effect
between the hole and electron pockets. Later on, it is found that the
K$_x$Fe$_{2-y}$Se$_2$ sample separates into two
phases\cite{shen2011intrinsic,li2012phase,zhang2011nodeless} - a
dominant antiferromagnetic insulating phase
K$_2$Fe$_4$Se$_5$\cite{ryan2011fe,wei2011novel,ye2011common}, and
a minority superconducting phase whose exact structure and compositions are still
under debate\cite{li2012phase,li2012kfe,texier2012nmr,
ding2013influence}. For the superconducting phase, it has been gradually conceived that the main structure is still the same as the typical BaFe$_2$As$_2$, but both the Fe and K may be deficient, and these deficiencies may form some kind of structures.  Recently, the neutron diffraction experiment indicates a potassium deficient but iron stoichiometric formula K$_x$Fe$_2$Se$_2$ for the superconducting phase\cite{carr2014}. Therefore the study of such material is complicated by the nature of mesoscopic phase separation\cite{wen2012overview,shen2011intrinsic,li2012phase,ryan2011fe,carr2014},
and it is very challenging to explore the properties of the normal
state of the superconducting phase. Recently, Yi \emph{et al}.
reported the study of ARPES on A$_x$Fe$_{2-y}$Se$_2$ (A = K, Rb) single crystals,
and proposed an orbital-selective Mott transition in the normal
state\cite{yi2013observation}. Moreover, femtosecond pump-probe
spectroscopy\cite{li2014mott} and THz
spectroscopy\cite{wang2014orbital} studies also evidenced the
Mott-transition-like behavior in the normal state. These
properties are consistent with the prediction of a multiband
theory assuming strong on-site Coulomb
interactions\cite{yu2013orbital}. Due to the reactivity of potassium
element, the sample of K$_x$Fe$_{2-y}$Se$_2$ is very sensitive in
air, therefore, as far as we know, the Hall effect has been seldom
investigated in K$_x$Fe$_{2-y}$Se$_2$. Previous preliminary
measurements reveal that the Hall coefficient is negative over the
whole temperature range, indicating that the system is dominated
by electronic-like charge carriers\cite{guo2010superconductivity,
fang2011fe,luo2011crystal}, being consistent with the observation
of the electron Fermi pockets in the ARPES measurements.

In this paper, we study the normal-state transport properties of
quenched K$_x$Fe$_{2-y}$Se$_2$ superconducting single crystals
with $T_c$ = 32 K through the in-plane resistivity, transverse
magnetoresistance (MR) and Hall effect measurements. We find that
the Kohler's rule is obeyed in the low temperature region. The
Hall coefficient has a strong but non-monotonic temperature
dependence below 150 K. This is in contrast with the FeAs-based
systems in which the Hall coefficient shows a monotonic
temperature dependence\cite{cheng2008,luo2009normal,fang2009roles}. These abnormal temperature-dependent
behaviors cannot be described by one single band model, suggesting
the multi-band nature in K$_x$Fe$_{2-y}$Se$_2$. Using a two-band
model (mainly $d_{xz/yz}$ and $d_{xy}$) analysis, we conclude that
a gap may open below 65 K, while the non-monotonic temperature
dependence of the Hall coefficient could be understood as a
consequence of an orbital-selective Mott transition. These results
would trigger further theoretical and experimental studies of the
orbital selective correlation effect in FeSCs.

\begin{figure}
\includegraphics[width=9cm]{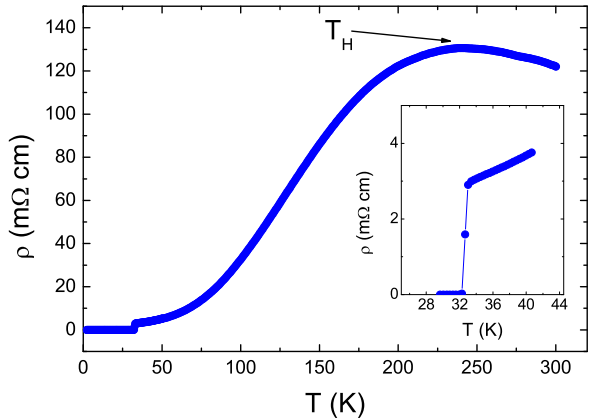}
\caption {(color online) Temperature dependence of in-plane
resistivity measured for the quenched K$_x$Fe$_{2-y}$Se$_2$ single
crystal with a broad hump appearing at $T_H \sim$ 240 K. A sharp
superconducting transition is observed at about 32 K. An enlarged view near the superconducting transition is shown as an inset. }\label{fig1}
\end{figure}

\begin{figure}
\includegraphics[width=9cm]{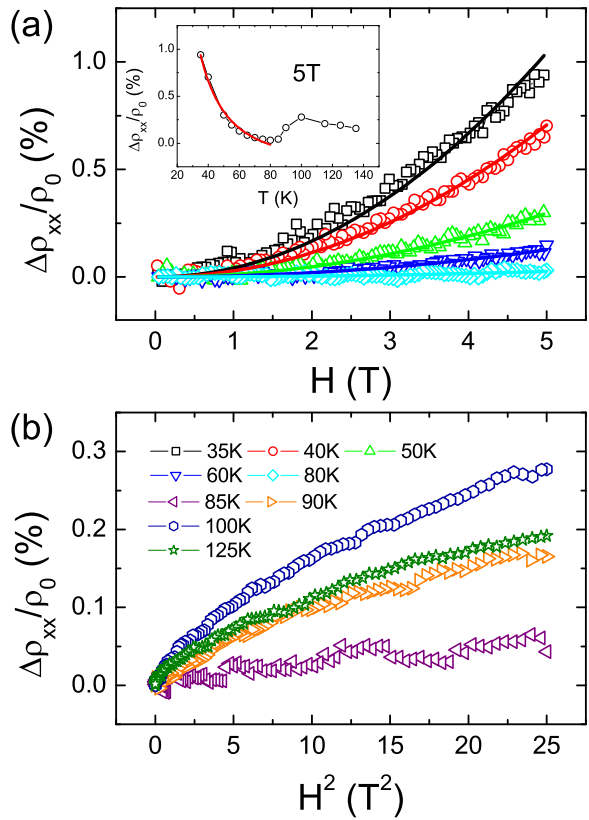}
\caption {Field dependence of the magnetoresistance of
K$_x$Fe$_{2-y}$Se$_2$. (a) The MR below 80 K. Solid lines are the
fits as a function of $H^2$. The inset shows temperature
dependence of MR measured at a magnetic field of 5 T. A pronounced
minimum occurs at 80 K. The red line is a fit to the data below 80
K as a function of $T^{-2}$. (b) The MR above 85 K plotted verse
$H^2$.} \label{fig2}
\end{figure}

\section{Measurements of magnetoresistance and Hall effect}

The K$_x$Fe$_{2-y}$Se$_2$ single crystals used for the transport
measurements were fabricated by a self-flux method with a starting synthesizing composition of K:Fe:Se=0.8:2:2. The crystals are rapidly quenched in liquid
nitrogen after heating to 350 $^{\circ}$C and staying for several hours. Details of the preparation were
described elsewhere\cite{ding2013influence}. The quenching process can greatly
improve the connection of the tiny superconducting networks (paths), and
thus the global appearance of superconductivity is much better
than the slowly cooled samples of
K$_x$Fe$_{2-y}$Se$_2$\cite{ding2013influence,han2012metastable,
liu2012evolution}. X-ray diffraction patterns (XRD) taken on the quenched crystals show only (00l) peaks with some small accompanying peaks. We think that the main diffraction (00l) peaks of the XRD pattern are coming from the major part of the sample, i.e., the K$_2$Fe$_4$Se$_5$ matrix, while the small accompanying peaks are coming from the minority superconducting networks. Because there is clear evidence of phase separation in the sample, it is meaningless to claim an uniform composition through out the sample. The microanalysis using energy-dispersive-spectrum (EDS) on the samples reveals that the background has a composition close to K:Fe:Se=2:4:5. For the present quenched sample, since the minority superconducting phase (path) has very small size which is smaller than the size of the electron beam in the EDS analysis, we could not use the EDS technique directly to get valid values of compositions for the three elements\cite{ding2013influence}. Transport
measurements were carried out with the six-lead method in a
Quantum Design instrument physical property measurement system
(PPMS). The electric contacts were made using silver
paste in a glove box filled with nitrogen atmosphere. We have worked on two samples from different batches and
the results are similar to each other.

In the K$_x$Fe$_{2-y}$Se$_2$ system, one concern is the
phase-separation property, the superconducting and insulating
phases could both contribute in transport measurements. As shown
in Fig. 1, a broad hump with the peak at $T_H \sim$ 240 K is
observed in the normal-state, being similar to those
reported previously\cite{guo2010superconductivity,ding2013influence,han2012metastable,
liu2012evolution}. This anomaly, being sensitive to the
preparation process, could be caused by the connection in series
between the metallic and the insulating phases as a result of the
phase-separation picture. Hence, we mainly focus on the
normal-state properties below $T_H$ which are dominated by
contributions from the metallic phase (superconducting paths below $T_c$). The small
residual resistivity with sharp superconducting transition ($T_c$
= 32 K) indicates good connectivity of the superconducting
paths.

In order to get more information of the normal-state properties,
further investigations are provided by the transverse MR and Hall
effect measurements on the same sample. In general, for most normal
metals, the MR exhibits a $H^2$-dependence in the weak-field limit,
and the MR normally affords a useful method to investigate the nature of
electronic scattering. In Fig. 2(a), we show the field dependence
of the transverse MR, $\Delta\rho_{xx}/\rho_0$, below 80 K, solid
lines are fits to $\Delta\rho_{xx}/\rho_0 = aH^2$, $\rho_0$ is the
resistivity under zero magnetic field. It is clear that the MR
increases as $H^2$ in the sweeping magnetic field up to 5 T,
signaling metallic behavior for \emph{T} $\leq$ 80 K, while the
slight deviation at 35 K can be attributed to the superconducting
fluctuations at a finite magnetic field. Besides, there is a
fundamental difference in the MR between \emph{T} $\leq$ 80 K and
T $\geq$ 85 K. In the latter, as shown in Fig. 2(b), the MR seems
not follow the $H^2$-dependence, which indicates a transition
of the electronic characteristics from one to another. The inset
of Fig. 2(a) shows the temperature dependence of MR at 5 T , a
clear minimum occurs at about 80 K and the MR below 80 K
shows a nice fit to the $T^{-2}$-dependence. It is interesting to note that, the crossover at around 80 K is consistent with the measurement of the finite-frequency dielectric function by means of terahertz spectroscopy in a Rb-based sister compound\cite{CharnukhaPRB}.

The Kohler's rule\cite{pippard1989magnetoresistance}, which
assumes a simple scaling function of $\Delta\rho/\rho_0 =
F(H/\rho_0)$, should be satisfied for a single band metal with an
isotropic Fermi surface, with $\rho_0$ the resistivity at a fixed
temperature and zero field. For a multiband system, this rule is
also applicable as long as the number of
charge carriers from each band is independent on temperature and the scattering rates of different bands
have the similar temperature dependence\cite{chan2014validity}.
From the first glance at the data below 60 K, as shown in Fig.
3(a), it seems that the Kohler's rule is slightly violated.
Actually this slight "deviation" of Kohler's rule may not be
true. The reason is that the resistivity at zero field, namely
$\rho_0(T)$=$\rho_0(T=0)+A/\tau$ does not really reflect directly
the scattering rate $1/\tau$ when the residual resistivity is
sizable. In addition, as argued below, there maybe a partial gap openning below 65 K, which may give an influence of the Kohler's scaling rule. Instead of using the original scaling function $\Delta\rho/\rho_0 = F(H/\rho_0)$, we use here a more accurate form of the
Kohler's scaling rule $\Delta\rho/\rho_0 = F(H\tau)$ where $\tau$ is
the relaxation rate\cite{luo2002Kohler}. Since the system exhibits
metallic properties in the low temperature region, we could assume
the relaxation rate as $\tau \propto T^{-2}$. Fig. 3(b) shows the
refined Kohler's plot of $\Delta\rho/\rho_0$ vs $(H\tau)^2 \propto
(HT^{-2})^2$. One can see that the Kohler's rule is well obeyed
below 80 K if we assume a general scattering rate $1/T^2$. In
contrast, the Kohler's rule is drastically violated above 85 K, as
shown with the enlarged view in the inset of Fig. 3(a).

We now switch our attention to the temperature dependence of Hall
coefficient. In the present K$_x$Fe$_{2-y}$Se$_2$ system, Hall
effect measurements may provide the message concerning the
temperature dependence of the charge carrier density and
mobilities of electrons in different bands of the
superconducting phase. Since the insulating phase
K$_2$Fe$_4$Se$_5$ has the nearest band 300 meV below the
Fermi energy, they should not contribute in the Hall effect
measurements below T$_H$. In Fig. 4(a), we show the Hall resistivity
$\rho_{xy}$ versus magnetic field up to 5 T, a linear relation
between $\rho_{xy}$ and magnetic field $H$ has been found in wide
temperature region (35 K to 150 K). From the $\rho_{xy}(H)$ data,
the Hall coefficient $R_H$ is determined through $R_H=\rho_{xy}/H$
and shown in Fig. 4(b). The negative $R_H$ over the whole
temperature region up to 150 K reveals that the conduction is
dominated by electron-like charge carriers. However, the most
remarkable feature in Fig. 4(b) is that the $R_H(T)$ shows a
strong but non-monotonic temperature dependence. This is in sharp
contrast with the FeAs-based 122 samples in which the Hall
coefficient is monotonically dependent on
temperature\cite{cheng2008,luo2009normal,fang2009roles}. By
having a closer scrutiny to the temperature dependence of Hall
coefficient, two characteristic temperatures could be defined:
$T_{gap}$ = 65 K and $T_{mott}$ = 85 K. Below 65 K, $R_H$
decreases rapidly with a suppression towards lower temperatures.
This is quite similar to that in the FeAs-based superconductors.
Between 65 K and 85 K, $R_H$ is almost temperature independent,
while it decreases upon raising temperature from 85 K to 150 K.
This anomalous behavior has never been reported in previous
studies in K$_x$Fe$_{2-y}$Se$_2$ and suggests that something
beyond the multi-band physics is very important here in
determining the electric conduction.

\section{Discussion}

\begin{figure}
\includegraphics[width=9cm]{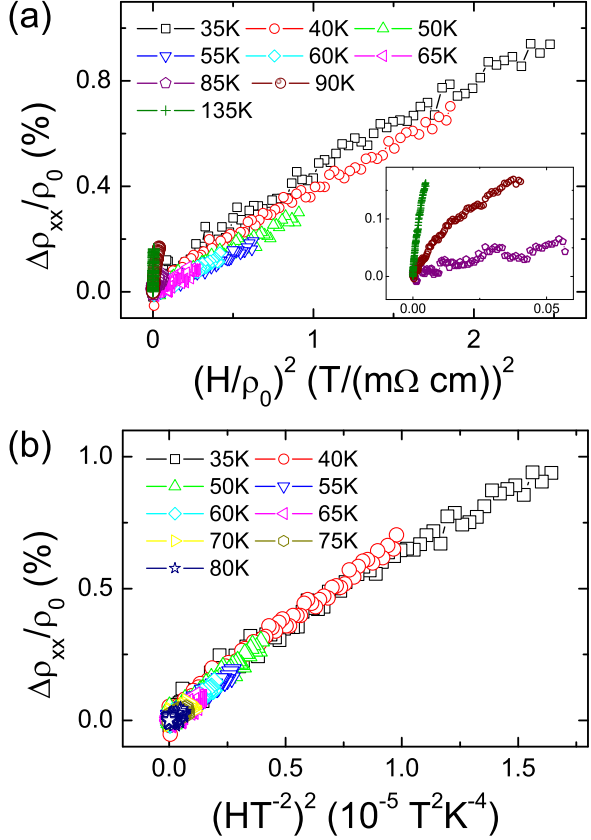}
\caption {(color online) (a) The scaling of Kohler's rule using
$\Delta\rho/\rho_0 = F(H/\rho_0)$ at various denoted temperatures.
Inset shows the Kohler's scaling of MR in low field region for
selective temperatures. (b) A refined Kohler's plot below 80 K
using $\Delta\rho/\rho_0 = F(H\tau)$ and assuming $1/ \tau \propto T^2$.}\label{fig3}
\end{figure}

\begin{figure}
\includegraphics[width=9cm]{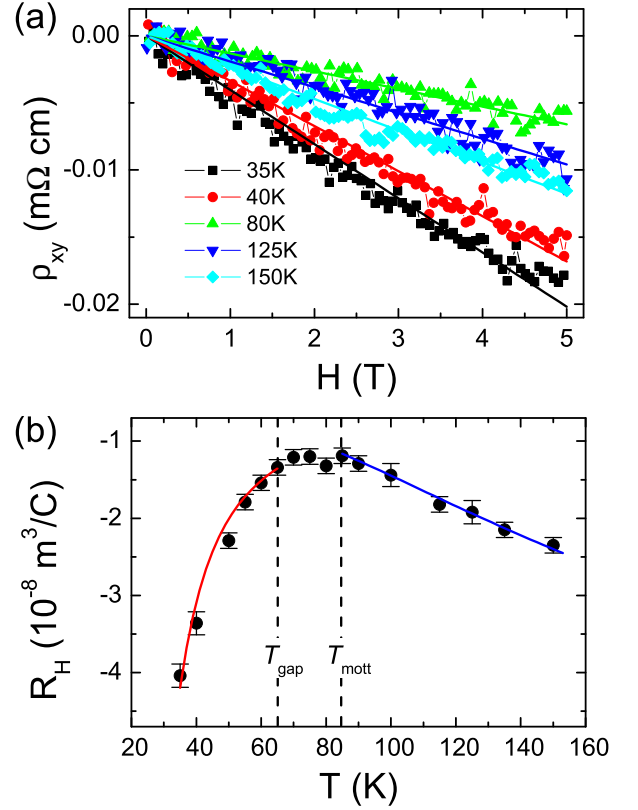}
\caption {(color online) (a) The transverse resistivity versus
the magnetic field at different temperatures. Solid lines are
linear fits. (b) Temperature dependence of Hall coefficient $R_H$,
two characteristic temperatures could be defined: $T_{gap}$ = 65 K and
$T_{mott}$ = 85 K. Solid lines are theoretical curves using the
two-band model (see text).}\label{fig4}
\end{figure}

According to the band structure calculations and ARPES studies,
two sets of electronic orbitals near the Fermi level,
namely $d_{xy}$ and $d_{xz/yz}$ play an important role. Since the $d_{xz}$ and $d_{yz}$ are
normally degenerate, we thus use a two band model to handle the
issue. Based on the Boltzmann transport theory in the weak field
limit, the equation of the Hall coefficient for two-band model
could be simplified as \begin{equation} R_H=\frac{\sigma^2_1R_{H1}
+ \sigma^2_2R_{H2}}{(\sigma_1 + \sigma_2)^2} = \frac{n_1\mu^2_1 +
n_2\mu^2_2}{-e(n_1\mu_1 + n_2\mu_2)^2}\end{equation} where
$\sigma_1 = en_1\mu_1, \sigma_2 = en_2\mu_2$ and $R_{H1} =
-1/en_1, R_{H2} = -1/en_2$ are single band conductivity and Hall
coefficients for the two orbitals, respectively. Based on the MR
analysis, the system exhibits metallic state below 80 K and a
$T^{-2}$ temperature dependence was found in our data. Thus, the
mobility of the two orbitals are expressed generally as $\mu_i =
\alpha_1T^{-2} (i = 1,2)$, $\alpha_1$ is a parameter related to
the effective density of states. Since the Kohler's rule is well
obeyed below 80 K, concerning the fact that the recent THz spectroscopy
experiments on Rb$_{1-x}$Fe$_{2-y}$Se$_2$ report a gap-like
suppression of optical conductance below 61 K\cite{wang2014orbital},
we thus attribute the temperature dependence of the Hall coefficient
below 65 K to the temperature dependent of the charge carrier density. In this scenario we then define the
carrier densities as $n_1 = n_1^0 \exp(-\Delta/k_BT)$ and $n_2 =
n_2^0 \exp(-\Delta/k_BT)$ where $\Delta$ is a partial gap which
opens at 65 K. The red line in Fig. 4(b) presents excellent
fitting results with the two-band model. One may suggest that the
temperature dependence of the Hall coefficient $R_H$ below 65 K is related to the
different temperature dependence of the scattering rate $1/\tau_i
(i = 1,2)$, this however cannot get support from other experiments,
for example the ARPES and optical data.

After crossing a maximum, $R_H$ drops with increasing temperature above 85 K,
which signals the involvement of a distinct electron scattering
mechanism. This turning point and the change of the temperature dependence is consistent with the results of MR
analysis. Recently, an orbital-selective Mott transition was
observed by ARPES at about 90 K in the superconducting phase of
K$_x$Fe$_{2-y}$Se$_2$, where the spectral weight near Fermi
surface for the $d_{xy}$ orbital diminishes while the other
orbitals $d_{xz/yz}$ remain metallic\cite{yi2013observation}.
Moreover, investigations using pump-probe spectroscopy\cite{li2014mott} and
THz spectroscopy\cite{wang2014orbital} also evidenced the
Mott-transition related behavior in the normal state. This
scenario could give a reasonable explanation to our data here. Since the
$d_{xz/yz}$ orbital remains metallic, we still express the
mobility of $d_{xz/yz}$ orbital as $\mu_1 = \alpha_1T^{-2}$ and
the carrier density as a constant $n_1^0$. Meanwhile, we consider
that the $d_{xy}$ band goes into the Mott phase with raising
temperature. As it is well known, the Mott insulating behavior has
been experimentally identified and theoretically explained in
terms of the band narrowing effect associated with the
electron-electron correlation. Therefore, the mobility of $d_{xy}$
orbital could be interpolated with the formula $\mu_2 =
\alpha_2T^{-\beta}/(1 + \gamma T)$, where $1/(1 + \gamma T)$ is
the modification term associated with the Mott transition. To
approach a solution, we may set the carrier density of the
$d_{xy}$ orbital as a constant $n'_2$ for simplicity. Thus, the
expression of the Hall coefficient with the orbital-selective Mott
phase is written as

\begin{equation} R_H = \frac{n_1^0(\alpha T^{-2})^2 +
n'_2(\frac{T^{-\beta}}{1 + \gamma T})^2}{-e(n_1^0\alpha T^{-2} +
n'_2(\frac{T^{-\beta}}{1 + \gamma T}))^2}\end{equation}

where $\alpha = \alpha_1/\alpha_2$ is the relative ratio of the
mobility coefficient of the two orbitals. In Fig. 4(b), the blue
solid line above 85 K shows the theoretical fitting result. Consequently, we acquire the
parameters as $\Delta$ = 7.7 meV, $n_1^0 = 9 \times 10^{26}$
m$^{-3}$,  $n_2^0 = 9 \times 10^{26}$ m$^{-3}$,  $n'_2 = 1.1
\times 10^{26}$ m$^{-3}$, $\alpha = 500, \beta = 0.2$ and $\gamma$
= 0.001 K$^{-1}$. The plateau of the Hall coefficient R$_H$ between 65 K and 85 K may be viewed
as the crossover of the two different regions.

We note that the appearance of a gap with the value of $\Delta$ =
7.7 meV below 65 K could be compared to the observation of
high-temperature superconductivity at 65 K in single-layer FeSe
films\cite{liu2012electronic,he2013phase}. We also can't rule out
the possibility of a pseudogap opening or other explanations for
this gap-like suppression of $R_H$ below 65 K. The decrease of $R_H$ for
increasing temperature starting from 85 K can get a strong support
from the scenario of the orbital-selective Mott transition in this
system. The explanation based on the orbital selective Mott transition should call for further theoretical and experimental efforts.

\section{Conclusion}

In summary, magnetoresistance and Hall coefficient $R_H$
have been measured in superconducting K$_x$Fe$_{2-y}$Se$_2$ single crystals.
The Kohler's rule is well obeyed below 80 K by assuming a general scattering rate $1/\tau \propto T^2$. We have observed a
strong and non-monotonic temperature dependence of the Hall
coefficient in the normal state. Using a two-band model analysis
and combining with the published data of the time domain optical conductivity
measurements, we conclude that a gap may open below 65 K, while
the data above 85 K could be understood as a consequence of an
orbital-selective Mott transition of the $d_{xy}$ band.

\begin{acknowledgments}
We appreciate the useful discussions with Qimiao Si and the help of Baigen Wang. This
work is supported by the NSF of China (11034011/A0402), the
Ministry of Science and Technology of China (973 projects:
2011CBA00102 and 2012CB821403) and PAPD.
\end{acknowledgments}

$^{\star}$ hhwen@nju.edu.cn

\end{document}